\RequirePackage{fix-cm}

\documentclass[%
floatfix,
showkeys,
nofootinbib, %
superscriptaddress, %
]{revtex4-1}

\usepackage{cmap}

\usepackage{ucs}
\usepackage[utf8x]{inputenc}
\usepackage[T1,T2A]{fontenc}
\usepackage[german,russian,english]{babel}

\usepackage[sort&compress]{natbib}
\usepackage{amsmath}
\usepackage{amssymb}

\usepackage{mathtools}
\mathtoolsset{
showonlyrefs,
mathic = true
}

\allowdisplaybreaks

\usepackage{hyperref}
\hypersetup{backref,
 colorlinks=false}
\hypersetup{pdfborder=0 0 0}

\usepackage{breakurl}

\usepackage{microtype}
\UseMicrotypeSet[protrusion]{alltext}

\usepackage{graphicx}

\graphicspath{{image/gen-sdu/en/}{image/gen-sdu/ru/}{image/gen-sdu/en/}{image/gen-sdu/}}

\usepackage[scanall]{psfrag}

\usepackage{fixltx2e}

\usepackage{nicefrac}

\makeatletter
\def\ps@pprintTitle{%
     \let\@oddhead\@empty
     \let\@evenhead\@empty
     \let\@oddfoot\@empty
     \let\@evenfoot\@oddfoot}
\makeatother

\usepackage{physics}
\usepackage{tensor}

\newcommand{\crd}[1]{\underline{\vphantom{j}{#1}}\,}

\newcommand{\setR}{\mathbb{R}}
\newcommand{\setN}{\mathbb{N}}

\begin{document}

\title{Two formalisms of stochastization of one-step models}

\author{D. S. Kulyabov}
\email{kulyabov_ds@rudn.university}
\affiliation{Department of Applied Probability and Informatics,\\
  Peoples' Friendship University of Russia (RUDN University),\\
  6 Miklukho-Maklaya St, Moscow, 117198, Russian Federation}
\affiliation{Laboratory of Information Technologies\\
  Joint Institute for Nuclear Research\\
  6 Joliot-Curie, Dubna, Moscow region, 141980, Russia}

\author{A. V. Korolkova}
\email{korolkova_av@rudn.university}
\affiliation{Department of Applied Probability and Informatics,\\
  Peoples' Friendship University of Russia (RUDN University),\\
  6 Miklukho-Maklaya St, Moscow, 117198, Russian Federation}

\author{L. A. Sevastianov}
\email{sevastianov_la@rudn.university}
\affiliation{Department of Applied Probability and Informatics,\\
  Peoples' Friendship University of Russia (RUDN University),\\
  6 Miklukho-Maklaya St, Moscow, 117198, Russian Federation}
\affiliation{Bogoliubov Laboratory of Theoretical Physics\\
  Joint Institute for Nuclear Research\\
  6 Joliot-Curie, Dubna, Moscow region, 141980, Russia}

\begin{abstract}
To construct realistic mathematical models from the first principles,
the authors suggest using the stochastization method. In a number of
works different approaches to stochastization of mathematical models
were considered. In the end, the whole variety of approaches was
reduced to two formalisms: combinatorial (state vectors) and operator
(occupation numbers). In the article the authors briefly describe
these formalisms with an emphasis on their practical application.
\end{abstract}

\maketitle

\section{Introduction}

The authors' appeal to stochastic differential equations was caused,
first of all, by the desire to build mathematical models from the
first principles. It turned out that many deterministic models (e.g.,
population models, network traffic
models~\cite{kulyabov:2014:icumt-2014:p2p, kulyabov:2014:icumt-2014:gns3}),
usually obtained ad-hoc, can be represented as deterministic part of
stochastic equations. These stochastic equations, in turn, are
obtained from the first principles.
In addition, in our opinion, the stochastic approach makes
mathematical models more realistic. For example, in the Lotka--Volterra
model~\cite{Feller:1939:acta_biotheoretica}, the atto-fox
problem~\cite{mollison:1991:atto-fox} is overcome.

For stochastization, the authors use two representations:
representation of state vectors (combinatorial approach) and
representation of occupation numbers (operator
approach)~\cite{kulyabov:2016:ecms:one-step, kulyabov:2016:epj:doi,
  grassberger:1980:fock-space, tauber:2005, janssen:2004, mobilia:2005}.

In the combinatorial approach, all operations are performed in the space
of states of the system, so we deal with a particular system
throughout manipulations with the model.
For the operator approach we can abstract from the specific implementation of
the system under study. We are working with abstract operators. We return to the
state space only at the end of the calculations. In addition,
we choose a particular operator algebra on the basis of symmetry of
the problem.

To construct stochastic models, we use the apparatus of interaction
schemes. Based on the interaction schemes, we have constructed a
diagram formalism for the operator approach.

In this article, the authors tried to give the main points of the
combinatorial and operator approaches in such a way that they could
be easily used in practical problems.

The structure of the article is as follows. In the section~\ref{sec:notation}
basic notations and conventions are introduced. 
Then the interaction
schemes are presented in the next
section~\ref{sec:schema}.
The combinatorial method of modelling is discussed in the
following section~\ref{sec:combinatorial}.
The operator model approach is presented in the
section~\ref{sec:operatorial}. 
Application of this technique is described in
section~\ref{sec:model} on the example of Verhulst model.

\section{Notations and conventions}
\label{sec:notation}

\begin{enumerate}

\item The abstract indices notation~(see \cite{penrose-rindler:spinors::en})
  is used in this work.  Under this notation a tensor as a whole
  object is denoted just as an index (e.g., $x^{i}$), components are
  denoted by underlined index (e.g., $x^{\crd{i}}$).

 \item We will adhere to the following agreements. Latin indices from
   the middle of the alphabet ($i$, $j$, $k$) will be applied to the space
   of the system state vectors. Latin indices from the beginning of
   the alphabet ($a$) will be related to the Wiener process space. Greek
   indices ($\alpha$) will set a number of different interactions in
   kinetic equations.

\end{enumerate}

\section{Interaction schemes}
\label{sec:schema}

Schemes of interaction are similar to the schemes of chemical
kinetics~\cite{waage:1986,gorban:2015}, which however have somewhat
different semantics.

The system state is defined by the vector
$\varphi^{i} \in \setR^n$, where $n$ is system dimension. The operator
$I^{i}_{j} \in \setN^{n}_{0} \times \setN^{n}_{0}$ describes the state 
of the system before the interaction, the operator
$F^{i}_{j} \in \setN^{n}_{0} \times \setN^{n}_{0}$ describes the state 
of the system
after the interaction. The result of interaction 
is the system transition from one state to another one:
\begin{equation}
  \label{eq:chemkin:simple}
  I^{i}_{j} \varphi^j
  \xrightleftharpoons[{\tensor*[^{-}]{k}{}}]{\tensor*[^{+}]{k}{}}
  F^{i}_{j} \varphi^{j}.
\end{equation}
The coefficients $\tensor*[^{+}]{k}{}$ and
$\tensor*[^{-}]{k}{}$ are the intensities of interaction.

There are $s$ types of interaction in our system, so instead of 
 $I^{i}_{j}$ and $F^{i}_{j}$ operators we will use operators
$I^{i \alpha}_{j} \in \setN^{n}_{0} \times \setN^{n}_{0} \times
\setN^{s}_{+}$
and
$F^{i \alpha}_{j} \in \setN^{n}_{0} \times \setN^{n}_{0} \times
\setN^{s}_{+}$:
\begin{equation}
  \label{eq:chemkin}
  I^{i \crd{\alpha}}_{j} \varphi^j
  \xrightleftharpoons[{\tensor*[^{-}]{k}{_{\crd{\alpha}}}}]{\tensor*[^{+}]{k}{_{\crd{\alpha}}}}
  F^{i \crd{\alpha}}_{j} \varphi^{j},
  \qquad \crd{\alpha} = \overline{1,s},
\end{equation}
the Greek indices specify the number of interactions and the Latin indices correspond to
the system order.

Usually, the diagonal operators $I$ and $F$ are used. That is,
they can be formally represented as a vector for each reaction. Then
the system~\eqref{eq:chemkin} takes the following form:
\begin{equation}
  \label{eq:chemkin2}
  I^{\crd{\alpha}}_{j} \varphi^j
  \xrightleftharpoons[{\tensor*[^{-}]{k}{_{\crd{\alpha}}}}]{\tensor*[^{+}]{k}{_{\crd{\alpha}}}}
  F^{\crd{\alpha}}_{j} \varphi^{j}.
\end{equation}
Here the following notation is used:
\begin{equation}
  \label{eq:n^i-notion}
  I^{\crd{\alpha}}_{j} := I^{i \crd{\alpha}}_{j} \delta_{i}, 
  \quad F^{\crd{\alpha}}_{j} := F^{i \crd{\alpha}}_{j} \delta_{i},
\end{equation}
where $\delta_{\crd{i}} = (1,\ldots,1)$.

In the combinatorial approach, the state transition operator is 
important:
\begin{equation}
  \label{eq:r_i}
  r_j^{i \crd{\alpha}} = F_j^{i \crd{\alpha}} -I_j^{i \crd{\alpha}},
  \quad
  r^{\crd{\alpha}}_{j} := r^{i \crd{\alpha}}_{j} \delta_{i}
  = F_j^{\crd{\alpha}} -I_j^{\crd{\alpha}}.
\end{equation}

\section{Combinatorial approach}
\label{sec:combinatorial}

In the combinatorial approach
for the system description we will use the master
equation, which
describes the transition probabilities for Markov
process~(see \cite{van-kampen:stochastic::en, gardiner:stochastic::en}).
\begin{equation}
  \label{eq:master:trans}
  \pdv{p(\varphi_{2},t_{2}|\varphi_{1},t_{1})}{t} = \int \bigl[
  w(\varphi_{2}|\psi,t_{2}) p(\psi,t_{2}|\varphi_{1},t_{1}) 
  -
  w(\psi|\varphi_{2},t_{2}) p(\varphi_{2},t_{2}|\varphi_{1},t_{1}) 
  \bigr] \dd{\psi},
\end{equation}
where $w(\varphi|\psi,t)$ is the probability of transition from the state $\psi$
to the state $\varphi$ per unit time.

By fixing the initial values of $\varphi_{1},t_{1}$, we can write the
equation for subensemble:
\begin{equation}
  \label{eq:master:subansemble}
  \pdv{p(\varphi,t)}{t} = \int
  \qty[
  w(\varphi|\psi,t) p(\psi,t) -
  w(\psi|\varphi,t) p(\varphi,t)
  ] \dd{\psi}.
\end{equation}

For the discrete domain of $\varphi$,
the~\eqref{eq:master:subansemble} can be written as follows
(the states are numbered by $n$ and $m$):
\begin{equation} 
  \label{eq:mas_eq}
    \pdv{p_{n}(t)}{t} = \sum\limits_{m} 
    \qty[w_{nm} p_{m}(t) - w_{mn} p_{n}(t)],
\end{equation}
where the $p_{n}$ is the probability of the system to be in a state $n$ at time $t$,
$w_{nm}$ is the probability of transition from the state $m$ into  the state $n$ per unit time.

There are two types of system transition from one state to another (based on one--step processes) 
as a result of system elements interaction: in the forward direction 
($\varphi^{i} + r^{i \crd{\alpha}}$) with the probability
$\tensor*[^{+}]{s}{_{\crd{\alpha}}}(\varphi^{k})$ and in the opposite direction
($\varphi^{i} - r^{i \crd{\alpha}}$) with the probability
$\tensor*[^{-}]{s}{_{\crd{\alpha}}}(\varphi^{k})$. 
The matrix of transition probabilities has the form: 
\begin{equation}
  w_{\crd{\alpha}}(\varphi^{i}| \psi^{i} ,t) = \tensor*[^{+}]{s}{_{\crd{\alpha}}}
  \delta_{\varphi^{i},\psi^{i}+1} + \tensor*[^{-}]{s}{_{\crd{\alpha}}}
  \delta_{\varphi^{i}, \psi^{i}-1},
  \qquad \crd{\alpha} = \overline{1,s},
\end{equation}
where $\delta_{i,j}$ is Kronecker delta.

Thus, the general form of the master equation for the state vector $\varphi^{i}$, 
changing by steps with length $r^{i \crd{\alpha}}$, is:
\begin{multline} 
  \label{eq:one-step:master:full}
  \pdv{p(\varphi^{i} ,t)}{t} = 
  \sum_{\crd{\alpha}=1}^{s} 
  \left\{
    \tensor*[^{-}]{s}{_{\crd{\alpha}}} (\varphi^{i}+r^{i
      \crd{\alpha}},t) p(\varphi^{i}+r^{i \crd{\alpha}} ,t) 
    + {} \right. \\ \left. {} + 
    \tensor*[^{+}]{s}{_{\crd{\alpha}}} (\varphi^{i}-r^{i \crd{\alpha}}, t) p(\varphi^{i} - r^{i \crd{\alpha}},t)
    -
    \qty[ 
    \tensor*[^{+}]{s}{_{\crd{\alpha}}} (\varphi^{i}) + 
    \tensor*[^{-}]{s}{_{\crd{\alpha}}} (\varphi^{i})
    ]
    p(\varphi^{i},t)      
  \right\}.
\end{multline}

In the combinatorial approach we don't use the master equation
itself. Instead, we use the Fokker--Planck equation and the Langevin
equation.

The Fokker--Planck equation is a special case of the master equation and
can be regarded as its approximation.
We can get through the expansion of the master equation in a series
up to the second order.
To do this, one can use the so-called Kramers--Moyal expansion~\cite{gardiner:stochastic::en}.

The Fokker--Planck equation has the following form:
\begin{equation}
  \label{eq:fp_nD}
  \pdv{p(\varphi^k, t)}{t} = 
  - \pdv{}{\varphi^{i}} \left[ A^{i}(\varphi^{k}) p(\varphi^{k}, t)
  \right] 
  +
  \frac{1}{2} \pdv{}{\varphi^{i}}{\varphi^{j}} 
  \left[ B^{i j} (\varphi^k) p(\varphi^{k}, t)  \right],
\end{equation}

The Langevin equation which corresponds to the Fokker--Planck equation:
\begin{equation}
  \label{eq:langevin}
  \dd \varphi^{i} = a^{i} \dd{t} + b^i_{a} \dd{W^{a}},
\end{equation}
where $a^{i} := a^{i} (\varphi^k)$, $b^{i}_{a} := b^{i}_{a}
(\varphi^k)$, 
$\varphi^i \in \setR^n $ is the system state vector, $W^{a} \in \mathbb{R}^m$
is the $m$-dimensional Wiener process.
Latin indices from the middle of the alphabet will be applied to the system state vectors 
(the dimensionality of space is $n$), and Latin indices from the beginning of 
the alphabet denote the variables related to the Wiener process vector 
(the dimensionality of space is $m \leqslant n$).

The relation between the equations~\eqref{eq:fp_nD} and~\eqref{eq:langevin}
is expressed by the following relationships:
\begin{equation}
  \label{eq:k-langevin}
  A^{i} = a^{i}, \qquad B^{i j} = b^{i}_{a} b^{j a}.
\end{equation}

 We will obtain the function $\tensor*[^{+}]{s}{_{\crd{\alpha}}}$ and
$\tensor*[^{-}]{s}{_{\crd{\alpha}}}$ with use of combinatorial considerations.
The transition rates $\tensor*[^{+}]{s}{_{\crd{\alpha}}}$ and $\tensor*[^{-}]{s}{_{\crd{\alpha}}}$ 
are proportional to the number of ways of choosing the number of
arrangements of $\varphi^{\crd{i}}$ to $I^{\crd{i} \crd{\alpha}}$
(denoted as $A_{\varphi^{\crd{i}}}^{I^{\crd{i} \crd{\alpha}}}$) and to
$F^{\crd{i} \crd{\alpha}}$ (denoted as $A_{\varphi^{\crd{i}}}^{F^{\crd{i} \crd{\alpha}}}$) and defined by:
\begin{equation}
  \label{eq:s-pm}
\begin{gathered}
  \tensor*[^{+}]{s}{_{\crd{\alpha}}} =
  \tensor*[^{+}]{k}{_{\crd{\alpha}}} \prod_{\crd{i}=1}^{n} 
  A_{\varphi^{\crd{i}}}^{I^{\crd{i} \crd{\alpha}}} =
  \tensor*[^{+}]{k}{_{\crd{\alpha}}} \prod_{\crd{i}=1}^{n}
  \frac{\varphi^{\crd{i}}!}{(\varphi^{\crd{i}} - I^{\crd{i} \crd{\alpha}})!}, \\
  \tensor*[^{-}]{s}{_{\crd{\alpha}}} = 
  \tensor*[^{-}]{k}{_{\crd{\alpha}}} \prod_{\crd{i}=1}^{n} 
  A_{\varphi^{\crd{i}}}^{F^{\crd{i} \crd{\alpha}}} =
  \tensor*[^{-}]{k}{_{\crd{\alpha}}} \prod_{\crd{i}=1}^{n}
  \frac{\varphi^{\crd{i}}!}{(\varphi^{\crd{i}}-F^{\crd{i} \crd{\alpha}})!}.
\end{gathered}
\end{equation}

Since the Fokker--Planck equation is a consequence of the expansion
of the master equation and by discarding small terms, we
will make an appropriate replacement in the equation~\eqref{eq:s-pm}. Namely,
combinations of the type $\varphi(\varphi-1) \cdots (\varphi - (n-1))$
will be replaced by $(\varphi)^n$:
\begin{equation}
  \label{eq:s-pm:exp}
\begin{gathered}
  \tensor*[^{+}_{\text{fp}}]{s}{_{\crd{\alpha}}} =
  \tensor*[^{+}]{k}{_{\crd{\alpha}}} \prod_{\crd{i}=1}^{n}
  (\varphi^{\crd{i}})^{I^{\crd{i} \crd{\alpha}}}, \\
  \tensor*[^{-}_{\text{fp}}]{s}{_{\crd{\alpha}}} = 
  \tensor*[^{-}]{k}{_{\crd{\alpha}}} \prod_{\crd{i}=1}^{n}
  (\varphi^{\crd{i}})^{F^{\crd{i} \crd{\alpha}}}.
\end{gathered}
\end{equation}

Then for the Fokker--Planck equation~\eqref{eq:fp_nD} we may obtain
formulas for the coefficients:
\begin{equation} 
  \label{eq:fp_coeff}
  \begin{gathered}
    A^{i} := A^{i}(\varphi^{k}) = r^{i \crd{\alpha}} 
    \qty[
      \tensor*[^+_{\text{fp}}]{s}{_{\crd{\alpha}}} -
      \tensor*[^-_{\text{fp}}]{s}{_{\crd{\alpha}}} ], \\
    B^{i j} := B^{i j}(\varphi^{k}) = r^{i \crd{\alpha}} r^{j
      \crd{\alpha}} 
    \qty[ \tensor*[^+_{\text{fp}}]{s}{_{\crd{\alpha}}} -
      \tensor*[^-_{\text{fp}}]{s}{_{\crd{\alpha}}} ].
  \end{gathered}
\end{equation}

Using the relation~\eqref{eq:k-langevin}, we may obtain the
coefficients for the Langevin equation~\eqref{eq:langevin} from
the relations~\eqref{eq:fp_coeff}.

\section{Operator approach} 
\label{sec:operatorial}

  The method of application of the second quantization formalism 
  for the non-quantum systems (statistical, deterministic systems) was
  studied in a series of articles~(see \cite{Doi:1976:second_quantization,
    Doi:1976:stochastic_theory, grassberger:1980:fock-space,
    peliti:1985:path_integral_approach}).
  The Dirac notation is commonly used for occupation numbers
  representation recording.

  The transition to the space of occupation numbers is not a unitary
  transformation.  However, the algorithm of transition can be constructed.

  Let's write the master equation~\eqref{eq:mas_eq} in the occupation
  number representation.  We will consider a system that does not
  depend on the spatial variables. For simplicity, we consider the
  one-dimensional version.

  Let's denote in~\eqref{eq:mas_eq} the probability that there are $n$
  particles in our system as $\varphi_{n}$:
\begin{equation}
  \label{eq:phi_n}
  \varphi_{n} := p_{n} (\varphi, t).
\end{equation}

The vector space $\mathcal{H}$ consists of states of $\varphi$.

Depending on the structure of the model, we can introduce the 
probability-based or the moment-based inner products~\cite{grassberger:1980:fock-space}.
We introduce a scalar product, exclusive ($\braket{}{}_{\text{ex}}$)
and inclusive ($\braket{}{}_{\text{in}}$). Let $\ket{n}$ be the 
basis vectors.
\begin{gather}
  \label{eq:scalar_prod_ex}
  \braket{\varphi}{\psi}_{\text{ex}} = \sum_n n! p_{n}^{*} (\varphi) p^{n}(\psi); \\
  \label{eq:scalar_prod_in}
  \braket{\varphi}{\psi}_{\text{in}} = \sum_n \frac{1}{k!} n_{k}^{*} (\varphi)
  n^{k}(\psi).
\end{gather}
Here $n_k$ are factorial moments:
\begin{equation}
  \label{eq:factorial_moment}
  n_k (\varphi) = \expval{n(n-1) \cdots (n-k+1)} = \pdv[k]{z}
  G(z,\varphi) |_{z=1},
\end{equation}
$G$ is generating function:
\begin{equation}
  \label{eq:generating_function}
  G (z, \varphi) = \sum_{n} z^{n} p_{n}(\varphi).
\end{equation}

Let's introduce creation and annihilation operators:
\begin{equation}
  \label{eq:creat+annihil}
  \begin{gathered}
    \pi \ket{n} = \ket{n+1}, \\
    a \ket{n} = n \ket{n-1}
  \end{gathered}
\end{equation}
and commutation rule\footnote{In fact, $a\pi\ket{n} -
  \pi a \ket{n} = (n+1)\ket{n} - n \ket{n} = \ket{n}$.}:
\begin{equation}
  \label{eq:commutator}
  [a,\pi] = 1.
\end{equation}

If the form of scalar product is~\eqref{eq:scalar_prod_ex} then
from~\eqref{eq:commutator} follows that our system is
described by  the Bose--Einstein statistics.

In the occupation numbers formalism the master equation becomes the Liouville
equation:
\begin{equation}
  \label{eq:liuville}
  \pdv{}{t} \ket{\varphi(t)} =  L \ket{\varphi(t)}.
\end{equation}
The Liouville operator $L$ satisfies the relation:
\begin{equation}
  \label{eq:liuville=0}
  \bra{0} L = 0.
\end{equation}

We will describe our proposed diagram technique for the stochastization of
one-step processes.

\begin{figure*}
  \begin{minipage}{0.45\linewidth}
    \centering
    \includegraphics[width=0.7\linewidth]{reaction}
  \caption{Forward interaction}
    \label{fig:reaction}
  \end{minipage}
  \hfill
  \begin{minipage}{0.45\linewidth}
    \centering
    \includegraphics[width=0.7\linewidth]{reaction-rev}
    \caption{Backward interaction}
    \label{fig:reaction-rev}
  \end{minipage}
\end{figure*}

We will write the the scheme of interaction in the form of
diagrams. Each scheme~\eqref{eq:chemkin} or~\eqref{eq:chemkin2}
corresponds to a pair of diagrams (see fig.~\ref{fig:reaction} and
\ref{fig:reaction-rev}) for forward and backward interaction
respectively.
The
diagram consists of the following elements.

\begin{itemize}
\item Incoming lines (in the fig.~\ref{fig:reaction} are denoted by the
  solid line). These lines are directed to the line of
  interaction. These lines are marked with the number and type of interacting
  entities. You can write a single entity per a line or group
  them.
\item Outgoing lines (in the fig.~\ref{fig:reaction} are denoted by the
   solid line). These lines are directed from the line of
   interaction. These lines are marked with the number and type of interacting
  entities. You can write a single entity per a line or group
  them.
\item Line of interaction (in the fig.~\ref{fig:reaction} is
  denoted by the dotted line). The direction of time is denoted by the
  arrow. This line is marked by the coefficient of rate of the interaction.
\end{itemize}

Each line is attributed to a certain factor (depending on the
the approach chosen). The resulting expression is obtained by multiplying
these factors.

We obtain the Liouville operator when using interaction diagrams in the
operator approach. Let us assign the corresponding factor for each
line. The resulting term is obtained as the normal ordered product of
factors.

\begin{figure*}
  \begin{minipage}{0.45\linewidth}
    \centering
    \includegraphics[width=0.7\linewidth]{reaction-operator}
  \caption{Forward interaction}
    \label{fig:reaction-operator}
  \end{minipage}
  \hfill
  \begin{minipage}{0.45\linewidth}
    \centering
    \includegraphics[width=0.7\linewidth]{reaction-rev-operator}
    \caption{Backward interaction}
    \label{fig:reaction-rev-operator}
  \end{minipage}
\end{figure*}

We use the following factors for each type of line
(Fig.~\ref{fig:reaction-operator}.).

\begin{itemize}
\item Incoming line. This line corresponds to the disappearance of one
  entity from the system. Therefore, it corresponds to the
  annihilation operator $a$. It is clear that the line with combined
  capacity $I$ corresponds to the operator $a^{I}$.
\item Outgoing line. This line corresponds to the emergence of one
  entity in the system. Therefore, it corresponds to the creation operator
   $\pi$. It is clear that the line with combined
  capacity $F$ corresponds to the operator $\pi^{F}$.
\item Line of interaction. This line corresponds to the ratio of the
  interaction rate.
\end{itemize}

\begin{figure*}
  \begin{minipage}{0.45\linewidth}
    \centering
    \includegraphics[width=0.7\linewidth]{reaction-operator-ext}
  \caption{Forward interaction, extended notation}
    \label{fig:reaction-operator-ext}
  \end{minipage}
  \hfill
  \begin{minipage}{0.45\linewidth}
    \centering
    \includegraphics[width=0.7\linewidth]{reaction-rev-operator-ext}
    \caption{Backward interaction, extended notation}
    \label{fig:reaction-rev-operator-ext}
  \end{minipage}
\end{figure*}

That is, for the diagram~\ref{fig:reaction-operator} we obtain a factor
$\tensor*[^{+}]{k}{} \pi^{F} a^{I} $. However, this violates the
equation~\eqref{eq:liuville=0}. 
Redressing this, we have to subtract the number of entities that have
entered into interaction, multiplied by the rate of the
interaction. Then we get a following term  of the Liouville operator:
\begin{equation}
  \label{eq:diagram-factor}
  \tensor*[^{+}]{k}{} \pi^{F} a^{I}  - \tensor*[^{+}]{k}{} 
  \pi^{I} a^{I} =
  \tensor*[^{+}]{k}{} 
  \qty(\pi^{F}  - \pi^{I}) a^{I}.
\end{equation}

To backward interactions (fig.~\ref{fig:reaction-rev-operator}), we
use the same rules.

To account for the additional factor of~\eqref{eq:diagram-factor} we
will use the extended diagrams (fig.~\ref{fig:reaction-operator-ext} and
\ref{fig:reaction-rev-operator-ext}). 
Here, from the the normal ordered product of the numerators 
the normal product of the denominators is subtracted.

Thus, the following Liouville operator corresponds to the
  scheme~\eqref{eq:chemkin}:
\begin{equation}
  \label{eq:liuville:chemkin}
  L = \sum_{\crd{\alpha},\crd{i}} \biggl[
  \tensor*[^{+}]{k}{_{\crd{\alpha}}}
  \qty(
  (\pi_{\crd{i}})^{F^{\crd{i}\crd{\alpha}}} -
  (\pi_{\crd{i}})^{I^{\crd{i}\crd{\alpha}}} 
  ) (a_{\crd{i}})^{I^{\crd{i}\crd{\alpha}}}
  + 
  \tensor*[^{-}]{k}{_{\crd{\alpha}}}
  \qty(
  (\pi_{\crd{i}})^{I^{\crd{i}\crd{\alpha}}} -
  (\pi_{\crd{i}})^{F^{\crd{i}\crd{\alpha}}} 
  ) (a_{\crd{i}})^{F^{\crd{i}\crd{\alpha}}}
  \biggr].
\end{equation}

\section{Verhulst model}
\label{sec:model}

As a demonstration of the method, we will consider the Verhulst 
model~\cite{verhulst:1838, Feller:1939:acta_biotheoretica,
  feller:1949:theory_stochastic_processes}, which describes the limited 
  growth. The attractiveness of this model is that it is one-dimensional and non-linear. 
Initially, this model was written down as the differential equation:
\begin{equation}
    \dv{\varphi}{t}= \lambda \varphi - \beta \varphi - \gamma \varphi^{2},
\end{equation}
where $\lambda$ denotes the breeding rate factor, $\beta$ is the
extinction rate factor, $\gamma$ is the factor of population
reduction rate (usually the rivalry of individuals is
considered). The same notation as in the original
  model~\cite{verhulst:1838} is used.

The interaction scheme for the stochastic version of the model is:
\begin{equation}
  \label{eq:verhulst:chemkin}
  \begin{gathered}
    \varphi \overset{\lambda}{\underset{\gamma}{\rightleftharpoons}} 2\varphi ,\\
    \varphi \xrightarrow{\beta} 0.
  \end{gathered}
\end{equation}

The interaction schemes~\eqref{eq:verhulst:chemkin} match
diagrams~\ref{fig:verhulst-1}, \ref{fig:verhulst-1-rev}, and \ref{fig:verhulst-2}.

The first relation means that an individual who eats one unit of meal is immediately reproduced, 
and in the opposite direction is the rivalry between individuals. 
The second relation describes the death of an individual.

\begin{figure*}
  \begin{minipage}{0.3\linewidth}
    \centering
    \includegraphics[width=\linewidth]{verhulst-1}
  \caption{First forward interaction}
\label{fig:verhulst-1}
  \end{minipage}
  \hfill
  \begin{minipage}{0.3\linewidth}
    \centering
    \includegraphics[width=\linewidth]{verhulst-1-rev}
    \caption{First backward interaction}
    \label{fig:verhulst-1-rev}
  \end{minipage}
  \hfill
  \begin{minipage}{0.3\linewidth}
    \centering
    \includegraphics[width=0.8\linewidth]{verhulst-2}
    \caption{Second forward interaction}
    \label{fig:verhulst-2}
  \end{minipage}
\end{figure*}

\subsection{Combinatorial approach}

Let's define transition rates  within
the Verhults model as follows:
\begin{equation}
  \begin{gathered}
    \tensor*[^{+}]{s}{_{1}} (\varphi) = \lambda \varphi, \\
    \tensor*[^{-}]{s}{_{1}} (\varphi) = \gamma \varphi (\varphi - 1), \\
    \tensor*[^{+}]{s}{_{2}} (\varphi) = \beta \varphi.
  \end{gathered}
  \quad
  \begin{gathered}
    \tensor*[^{+}]{s}{_{1}} (\varphi -1) = \lambda (\varphi -1), \\
    \tensor*[^{-}]{s}{_{1}} (\varphi -1) = \gamma (\varphi -1) (\varphi - 2), \\
    \tensor*[^{+}]{s}{_{2}} (\varphi -1) = \beta (\varphi -1).
  \end{gathered}
  \quad
  \begin{gathered}
    \tensor*[^{+}]{s}{_{1}} (\varphi +1) = \lambda (\varphi +1), \\
    \tensor*[^{-}]{s}{_{1}} (\varphi +1) = \gamma (\varphi +1) \varphi, \\
    \tensor*[^{+}]{s}{_{2}} (\varphi +1) = \beta (\varphi +1).
  \end{gathered}
\end{equation}

\begin{equation}
  r^1 = 1, \qquad
  r^2 = -1.
\end{equation}

Then, based on~\eqref{eq:one-step:master:full}, the form of the master equation is:
\begin{equation}
  \label{eq:master}
  \pdv{p (\varphi,t)}{t} =  
  - \qty[\lambda \varphi + \beta \varphi +
  \gamma \varphi (\varphi -1)] 
  p(\varphi,t) 
  +
  \qty[\beta (\varphi + 1)
  +  \gamma (\varphi + 1) \varphi] 
  p(\varphi + 1,t) + \lambda (\varphi - 1) p (\varphi - 1,t).
\end{equation}

For particular values of $\varphi=n$ (see Eq.~\eqref{eq:mas_eq}):
\begin{multline}
  \label{eq:verhulst:master:n}
  \pdv{p_{n}(t)}{t} := \pdv{p (\varphi,t)}{t}\eval_{\varphi=n} 
  = {} \\ {} =
  - \qty[\lambda n + \beta n +
  \gamma n(n-1)] 
  p_{n}(t) 
  +
  \qty[\beta (n + 1) + \gamma (n + 1)n] 
  p_{n+1}(t) + 
  \lambda (n - 1) p_{n-1} (t).
\end{multline}

\subsection{Operator approach}

The interaction schemes~\eqref{eq:verhulst:chemkin} in the operator approach  match
diagrams~\ref{fig:verhulst-1-operator}, \ref{fig:verhulst-1-rev-operator}, and \ref{fig:verhulst-2-operator}.

From~\eqref{eq:verhulst:chemkin} and~\eqref{eq:liuville:chemkin} the 
Liouville operator is:
\begin{multline}
  \label{eq:verhulst:liuville}
  L = \lambda (\pi^2 -\pi) a + \gamma (\pi - \pi^2) a^2 + 
  \beta (1 - \pi) a 
  = {} \\ {} =
  \lambda \qty((a^{\dagger})^2 - a^{\dagger}) a + 
  \gamma \qty(a^{\dagger} - (a^{\dagger})^2) a^2 + 
  \beta \qty(1 - a^{\dagger}) a 
  = {} \\ {} =
  \lambda \qty(a^{\dagger} - 1)a^{\dagger} a +
  \beta \qty(1 - a^{\dagger}) a + 
  \gamma \qty(1 - a^{\dagger})a^{\dagger} a^2.
\end{multline}

\begin{figure*}
  \begin{minipage}{0.3\linewidth}
    \centering
    \includegraphics[width=\linewidth]{verhulst-1-operator}
  \caption{First forward interaction}
\label{fig:verhulst-1-operator}
  \end{minipage}
  \hfill
  \begin{minipage}{0.3\linewidth}
    \centering
    \includegraphics[width=\linewidth]{verhulst-1-rev-operator}
    \caption{First backward interaction}
    \label{fig:verhulst-1-rev-operator}
  \end{minipage}
  \hfill
  \begin{minipage}{0.3\linewidth}
    \centering
    \includegraphics[width=0.8\linewidth]{verhulst-2-operator}
    \caption{Second forward interaction}
    \label{fig:verhulst-2-operator}
  \end{minipage}
\end{figure*}

The master equation by the Liouville operator:
\begin{multline}
  \label{eq:verhulst:master2liuville}
  \pdv{p_{n}(t)}{t} = \frac{1}{n!} \bra{n}L\ket{\varphi} 
  =  {} \\ {} =
  \frac{1}{n!} \bra{n}
  - \qty[ 
  \lambda a^{\dagger} a +
  \beta a^{\dagger} a +
  \gamma a^{\dagger}a^{\dagger} a a
  ]
  +
  \qty[
  \beta a + 
  \gamma a^{\dagger} a a
  ]
  +
  \lambda a^{\dagger}a^{\dagger} a 
  \ket{\varphi}
  = {} \\ {} =
  - \qty[\lambda n + \beta n +
  \gamma n(n-1)] 
  \bra{n}\ket{\varphi}
  + {} \\ {} +
  \qty[\beta (n + 1) + \gamma (n + 1)n] 
  \bra{n+1}\ket{\varphi} + 
  \lambda (n - 1) \bra{n-1}\ket{\varphi}
  = {} \\ {} =
  - \qty[\lambda n + \beta n +
  \gamma n(n-1)] 
  p_{n}(t) 
  + {} \\ {} +
  \qty[\beta (n + 1) + \gamma (n + 1)n] 
  p_{n+1}(t) + 
  \lambda (n - 1) p_{n-1} (t).
\end{multline}

The result~\eqref{eq:verhulst:master2liuville} coincides with 
the formula~\eqref{eq:verhulst:master:n}, which was obtained by combinatorial method.

\section{Conclusions}
\label{sec:conclusion}

The described stochastization methods make it possible to obtain a
specific form of both the Liouville operator and self-consistent
stochastic differential equations in the form of Langevin and
Fokker--Planck. The authors hope that the proposed diagram technique
will simplify the derivation of the Liouville operator in the
representation of occupation numbers.

\begin{acknowledgments}

The publication has been prepared with the support of the ``RUDN University Program 5-100'' 
and funded by Russian Foundation for Basic Research (RFBR) according to the research projects 
No~16-07-00556 and 18-07-00567.

\end{acknowledgments}

 \bibliographystyle{elsarticle-num}

\bibliography{bib/gen-sdu/cite}

\end{document}


\title{Два формализма стохастизации одношаговых моделей}

\author{Д. С. Кулябов}
\email{kulyabov_ds@rudn.university}
\affiliation{Кафедра прикладной информатики и теории вероятностей,\\
  Российский университет дружбы народов,\\
  117198, Москва, ул. Миклухо-Маклая, д. 6}
\affiliation{Лаборатория информационных технологий,\\
  Объединённый институт ядерных исследований,\\
  ул. Жолио-Кюри 6, Дубна, Московская область, Россия, 141980}

\author{А. В. Королькова}
\email{korolkova_av@rudn.university}
\affiliation{Кафедра прикладной информатики и теории вероятностей,\\
  Российский университет дружбы народов,\\
  117198, Москва, ул. Миклухо-Маклая, д. 6}

\author{Л. А. Севастьянов}
\email{sevastianov_la@rudn.university}
\affiliation{Кафедра прикладной информатики и теории вероятностей,\\
  Российский университет дружбы народов,\\
  117198, Москва, ул. Миклухо-Маклая, д. 6}
\affiliation{Лаборатория теоретической физики,\\
  Объединённый институт ядерных исследований,\\
  ул. Жолио-Кюри 6, Дубна, Московская область, Россия, 141980}

\begin{abstract}

  Для построения реалистичных математических моделей из первых
  принципов авторы предлагают использовать метод стохастизации.
  В ряде работ были рассмотрены разные подходы к стохастизации
  математических моделей. В конечном счёте всё многообразие подходов
  было сведено к двум формализмам: комбинаторному (векторов состояний)
  и операторному (чисел заполнения). В статье авторы кратко описывают
  эти формализмы с упором на их практическое применение.

\end{abstract}

  \keywords{представление чисел заполнения, пространство Фока, нотация
  Дирака, одношаговые процессы,  основное кинетическое уравнение,
  диаграммная техника}

\maketitle

\section{Введение}

Обращение авторов к стохастическим дифференциальным уравнениям было
вызвано, в первую очередь, желанием строить математические модели из
первых принципов. Оказалось, что многие детерминистические модели
(например, популяционные модели, модели сетевого трафика~\cite{kulyabov:2014:icumt-2014:p2p,
  kulyabov:2014:icumt-2014:gns3}), обычно получаемые ad-hoc, можно
представить как детерминистическую часть стохастических уравнений. Эти
стохастические уравнения, в свою очередь, получаются из первых
принципов.
Кроме того, по нашему мнению, стохастический подход делает
математические модели более реалистичными. Например, в модели
Лотки--Вольтерры~\cite{Feller:1939:acta_biotheoretica} преодолевается атто-лисья проблема~\cite{mollison:1991:atto-fox}.

Для стохастизации авторы используют два представления: представление векторов состояния
(комбинаторный подход) и представление
чисел заполнения (операторный
подход)~\cite{kulyabov:2016:ecms:one-step, kulyabov:2016:epj:doi, grassberger:1980:fock-space,
  tauber:2005, janssen:2004, mobilia:2005}.

В комбинаторном подходе все действия выполняются в пространстве
векторов состояния системы, то есть мы на протяжении всех модельных
манипуляций имеем дело с конкретной исследуемой системой.
В операторном подходе мы отвлекаемся от конкретной реализации
исследуемой системы, работая с абстрактными операторами. В
пространство векторов состояний мы переходим только в конце
вычислений. Кроме того, конкретную операторную алгебру мы выбираем,
исходя их симметрии задачи.

Для конструирования стохастических моделей нами используется аппарат
схем взаимодействия. На основе схем взаимодействия нами построен
диаграммный формализм для операторного подхода.

В данной статье авторы постарались дать основные моменты
комбинаторного и операторного подходов таким образом, чтобы их можно
было легко использовать в практических задачах.

Структура статьи следующая. В разделе~\ref{sec:notation} введены
основные обозначения и соглашения.
В разделе~\ref{sec:schema} вводятся
схемы взамодействия. В разделе~\ref{sec:combinatorial} описывается
комбинаторный модельный подход. 
В разделе~\ref{sec:operatorial} даётся операторный модельный подход и
диаграммная техника.
В разделе~\ref{sec:model}
демонстрируется применение описанных подходов 
на примере модели Ферхюльста.

\section{Обозначения и соглашения}
\label{sec:notation}

\begin{enumerate}

\item В работе используется нотация абстрактных
  индексов~\cite{penrose-rindler:spinors::ru}. В данной нотации тензор как
  целостный объект обозначается просто индексом (например, $x^{i}$),
  компоненты обозначаются подчёркнутым индексом (например,
  $x^{\crd{i}}$).

\item Будем придерживаться следующих соглашений. Латинские индексы из
  середины алфавита ($i$, $j$, $k$) будут относиться к пространству
  векторов состояний системы. Латинские индексы из начала алфавита
  ($a$) будут относиться к пространству винеровского
  процесса. Греческие индексы ($\alpha$) будут задавать количество
  разных взаимодействий в кинетических уравнениях.

\end{enumerate}

\section{Схемы взаимодействия}
\label{sec:schema}

Схемы взаимодействия подобны схемам химической
кинетики~\cite{waage:1986, gorban:2015}, которые, впрочем, имеют
несколько другую семантику.

Состояние системы будем описывать вектором состояния
$\varphi^{i} \in \setR^n$, где $n$~--- размерность
системы. Оператор
$I^{i}_{j} \in \setN^{n}_{0} \times \setN^{n}_{0}$ задаёт
состояние
системы до взаимодействия, оператор
$F^{i}_{j} \in \setN^{n}_{0} \times \setN^{n}_{0}$~---
после.
В результате
взаимодействия происходит переход системы в другое состояние:
\begin{equation}
  \label{eq:chemkin:simple}
  I^{i}_{j} \varphi^j
  \xrightleftharpoons[{\tensor*[^{-}]{k}{}}]{\tensor*[^{+}]{k}{}}
  F^{i}_{j} \varphi^{j}.
\end{equation}
Коэффициенты $\tensor*[^{+}]{k}{}$ и
$\tensor*[^{-}]{k}{}$ имеют смысл интенсивностей взаимодействия.

В~системе может происходить $s$ видов различных
взаимодействий. Поэтому вместо операторов $I^{i}_{j}$ и $F^{i}_{j}$
будем рассматривать операторы
$I^{i \alpha}_{j} \in \setN^{n}_{0} \times \setN^{n}_{0} \times
\setN^{s}_{+}$
и
$F^{i \alpha}_{j} \in \setN^{n}_{0} \times \setN^{n}_{0} \times
\setN^{s}_{+}$:
\begin{equation}
  \label{eq:chemkin}
  I^{i \crd{\alpha}}_{j} \varphi^j
  \xrightleftharpoons[{\tensor*[^{-}]{k}{_{\crd{\alpha}}}}]{\tensor*[^{+}]{k}{_{\crd{\alpha}}}}
  F^{i \crd{\alpha}}_{j} \varphi^{j},
  \qquad \crd{\alpha} = \overline{1,s},
\end{equation}
здесь греческие индексы задают количество взаимодействий, а латинские~---
размерность системы.

Обычно используют диагональные операторы $I$ и $F$.
То есть формально их можно представить как вектор для каждой реакции.
Тогда система~\eqref{eq:chemkin} приобретает следующий вид:
\begin{equation}
  \label{eq:chemkin2}
  I^{\crd{\alpha}}_{j} \varphi^j
  \xrightleftharpoons[{\tensor*[^{-}]{k}{_{\crd{\alpha}}}}]{\tensor*[^{+}]{k}{_{\crd{\alpha}}}}
  F^{\crd{\alpha}}_{j} \varphi^{j}.
\end{equation}
Здесь использованы следующие обозначения:
\begin{equation}
  \label{eq:n^i-notion}
  I^{\crd{\alpha}}_{j} := I^{i \crd{\alpha}}_{j} \delta_{i}, 
  \quad F^{\crd{\alpha}}_{j} := F^{i \crd{\alpha}}_{j} \delta_{i},
\end{equation}
где $\delta_{\crd{i}} = (1,\ldots,1)$.

Для комбинаторного подхода также важен оператор  
изменения состояния:
\begin{equation}
  \label{eq:r_i}
  r_j^{i \crd{\alpha}} = F_j^{i \crd{\alpha}} -I_j^{i \crd{\alpha}},
  \quad
  r^{\crd{\alpha}}_{j} := r^{i \crd{\alpha}}_{j} \delta_{i}
  = F_j^{\crd{\alpha}} -I_j^{\crd{\alpha}}.
\end{equation}

\section{Комбинаторный подход}
\label{sec:combinatorial}

В комбинаторном подходе
для описания системы используется основное
кинетическое уравнение (\emph{master equation}), определяющее
вероятность перехода в соседнее состояние  для марковского
процесса~\cite{van-kampen:stochastic::ru, gardiner:stochastic::ru}.
\begin{equation}
  \label{eq:master:trans}
  \pdv{p(\varphi_{2},t_{2}|\varphi_{1},t_{1})}{t} = \int \bigl[
  w(\varphi_{2}|\psi,t_{2}) p(\psi,t_{2}|\varphi_{1},t_{1}) 
  -
  w(\psi|\varphi_{2},t_{2}) p(\varphi_{2},t_{2}|\varphi_{1},t_{1}) 
  \bigr] \dd{\psi},
\end{equation}
где $w(\varphi|\psi,t)$ есть вероятность перехода из состояния $\psi$
в состояние $\varphi$ за единицу времени.

Зафиксировав начальные значения $\varphi_{1},t_{1}$, можно записать 
данное уравнение для подансамбля:
\begin{equation}
  \label{eq:master:subansemble}
  \pdv{p(\varphi,t)}{t} = \int
  \qty[
  w(\varphi|\psi,t) p(\psi,t) -
  w(\psi|\varphi,t) p(\varphi,t)
  ] \dd{\psi}.
\end{equation}

При дискретной области определения множества состояний системы $\varphi$ можно
записать~\eqref{eq:master:subansemble} (пронумеровав состояния числами
$n$ и $m$):
\begin{equation} 
  \label{eq:mas_eq}
    \pdv{p_{n}(t)}{t} = \sum\limits_{m} 
    \qty[w_{nm} p_{m}(t) - w_{mn} p_{n}(t)],
\end{equation}
где $p_{n}$~--- вероятность нахождения
системы в состоянии $n$ в момент времени $t$,
$w_{nm}$~--- вероятность перехода системы из состояния $m$ в
состояние $n$ за единицу времени.

В системе, описываемой одношаговыми процессами, возможны два вида
перехода системы из одного состояния в другое, происходящие в
результате взаимодействия элементов в прямом направлении
($\varphi^{i} + r^{i \crd{\alpha}}$) с вероятностью
$\tensor*[^{+}]{s}{_{\crd{\alpha}}}(\varphi^{k})$ и в обратном
направлении ($\varphi^{i} - r^{i \crd{\alpha}}$) с
вероятностью
$\tensor*[^{-}]{s}{_{\crd{\alpha}}}(\varphi^{k})$. А
матрица вероятностей переходов может быть записана в виде:
\begin{equation}
  w_{\crd{\alpha}}(\varphi^{i}| \psi^{i} ,t) = \tensor*[^{+}]{s}{_{\crd{\alpha}}}
  \delta_{\varphi^{i},\psi^{i}+1} + \tensor*[^{-}]{s}{_{\crd{\alpha}}}
  \delta_{\varphi^{i}, \psi^{i}-1},
  \qquad \crd{\alpha} = \overline{1,s},
\end{equation}
где $\delta_{i,j}$ --- символ Кронекера.

Таким образом, общий вид основного кинетического уравнения для вектора
состояний $\varphi^{i}$, изменяющегося шагами длины $r^{i \crd{\alpha}}$, принимает вид:
\begin{multline} 
  \label{eq:one-step:master:full}
  \pdv{p(\varphi^{i} ,t)}{t} = 
  \sum_{\crd{\alpha}=1}^{s} 
  \left\{
    \tensor*[^{-}]{s}{_{\crd{\alpha}}} (\varphi^{i}+r^{i
      \crd{\alpha}},t) p(\varphi^{i}+r^{i \crd{\alpha}} ,t) 
    + {} \right. \\ \left. {} + 
    \tensor*[^{+}]{s}{_{\crd{\alpha}}} (\varphi^{i}-r^{i \crd{\alpha}}, t) p(\varphi^{i} - r^{i \crd{\alpha}},t)
    -
    \qty[ 
    \tensor*[^{+}]{s}{_{\crd{\alpha}}} (\varphi^{i}) + 
    \tensor*[^{-}]{s}{_{\crd{\alpha}}} (\varphi^{i})
    ]
    p(\varphi^{i},t)      
  \right\}.
\end{multline}

В комбинаторном подходе собственно основное кинетическое уравнение не
используется. Вместо него используют уравнение Фоккера--Планка и
уравнение Ланжевена.

Уравнение Фоккера--Планка является частным случаем основного
кинетического уравнения и может рассматриваться как его приближённая
форма. Его можно получить путём разложения основного кинетического
уравнения в ряд до членов второго порядка вкючительно.
Для этого можно использовать
так называемое
разложение
Крамерса--Мойала~\cite{gardiner:stochastic::ru}.

Уравнение Фоккера--Планка имеет следующий вид:
\begin{equation}
  \label{eq:fp_nD}
  \pdv{p(\varphi^k, t)}{t} = 
  - \pdv{}{\varphi^{i}} \left[ A^{i}(\varphi^{k}) p(\varphi^{k}, t)
  \right] 
  +
  \frac{1}{2} \pdv{}{\varphi^{i}}{\varphi^{j}} 
  \left[ B^{i j} (\varphi^k) p(\varphi^{k}, t)  \right],
\end{equation}

Уравнению Фоккера--Планка соответствует уравнение Ланжевена:
\begin{equation}
  \label{eq:langevin}
  \dd \varphi^{i} = a^{i} \dd{t} + b^i_{a} \dd{W^{a}},
\end{equation}
где $a^{i} := a^{i} (\varphi^k)$, $b^{i}_{a} := b^{i}_{a}
(\varphi^k)$, 
$\varphi^i \in \setR^n $ --- вектор состояний системы, $W^{a} \in \mathbb{R}^m$
--- $m$-мерный винеровский процесс.
Здесь латинскими индексами из середины алфавита обозначаются величины,
относящиеся к векторам состояний (размерность пространства $n$), а
латинскими индексами из начала алфавита обозначаются величины,
относящиеся к вектору винеровского процесса (размерность пространства
$m \leqslant n$).

При этом связь между уравнениями \eqref{eq:fp_nD} и \eqref{eq:langevin}
выражается следующими соотношениями:
\begin{equation}
  \label{eq:k-langevin}
  A^{i} = a^{i}, \qquad B^{i j} = b^{i}_{a} b^{j a}.
\end{equation}

Для получения $\tensor*[^{+}]{s}{_{\crd{\alpha}}}$ и
$\tensor*[^{-}]{s}{_{\crd{\alpha}}}$ используют комбинаторные соображения.
Интенсивности перехода в единицу времени
$\tensor*[^{+}]{s}{_{\crd{\alpha}}}$ и $\tensor*[^{-}]{s}{_{\crd{\alpha}}}$ пропорциональны соответственно числу
способов выбора числа размещений из $\varphi^{\crd{i}}$ по $I^{\crd{i}
  \crd{\alpha}}$ (обозначается как $A_{\varphi^{\crd{i}}}^{I^{\crd{i} \crd{\alpha}}}$) и по
$F^{\crd{i} \crd{\alpha}}$ (обозначается как $A_{\varphi^{\crd{i}}}^{F^{\crd{i} \crd{\alpha}}}$) и определяются выражениями:
\begin{equation}
  \label{eq:s-pm}
\begin{gathered}
  \tensor*[^{+}]{s}{_{\crd{\alpha}}} =
  \tensor*[^{+}]{k}{_{\crd{\alpha}}} \prod_{\crd{i}=1}^{n} 
  A_{\varphi^{\crd{i}}}^{I^{\crd{i} \crd{\alpha}}} =
  \tensor*[^{+}]{k}{_{\crd{\alpha}}} \prod_{\crd{i}=1}^{n}
  \frac{\varphi^{\crd{i}}!}{(\varphi^{\crd{i}} - I^{\crd{i} \crd{\alpha}})!}, \\
  \tensor*[^{-}]{s}{_{\crd{\alpha}}} = 
  \tensor*[^{-}]{k}{_{\crd{\alpha}}} \prod_{\crd{i}=1}^{n} 
  A_{\varphi^{\crd{i}}}^{F^{\crd{i} \crd{\alpha}}} =
  \tensor*[^{-}]{k}{_{\crd{\alpha}}} \prod_{\crd{i}=1}^{n}
  \frac{\varphi^{\crd{i}}!}{(\varphi^{\crd{i}}-F^{\crd{i} \crd{\alpha}})!}.
\end{gathered}
\end{equation}

Поскольку уравнение Фоккера--Планка является следствием разложения в
ряд основного кинетического уравнения и отбрасывание малых членов,
сделаем в~\eqref{eq:s-pm} соответствующую замену. А именно 
комбинации типа
$\varphi (\varphi-1) \cdots (\varphi - (n-1))$ заменим на на $(\varphi)^n$:
\begin{equation}
  \label{eq:s-pm:exp}
\begin{gathered}
  \tensor*[^{+}_{\text{fp}}]{s}{_{\crd{\alpha}}} =
  \tensor*[^{+}]{k}{_{\crd{\alpha}}} \prod_{\crd{i}=1}^{n}
  (\varphi^{\crd{i}})^{I^{\crd{i} \crd{\alpha}}}, \\
  \tensor*[^{-}_{\text{fp}}]{s}{_{\crd{\alpha}}} = 
  \tensor*[^{-}]{k}{_{\crd{\alpha}}} \prod_{\crd{i}=1}^{n}
  (\varphi^{\crd{i}})^{F^{\crd{i} \crd{\alpha}}}.
\end{gathered}
\end{equation}

Тогда для уравнения Фоккера--Планка~\eqref{eq:fp_nD} получим формулы для коэффициентов:
\begin{equation} 
  \label{eq:fp_coeff}
  \begin{gathered}
    A^{i} := A^{i}(\varphi^{k}) = r^{i \crd{\alpha}} 
    \qty[
      \tensor*[^+_{\text{fp}}]{s}{_{\crd{\alpha}}} -
      \tensor*[^-_{\text{fp}}]{s}{_{\crd{\alpha}}} ], \\
    B^{i j} := B^{i j}(\varphi^{k}) = r^{i \crd{\alpha}} r^{j
      \crd{\alpha}} 
    \qty[ \tensor*[^+_{\text{fp}}]{s}{_{\crd{\alpha}}} -
      \tensor*[^-_{\text{fp}}]{s}{_{\crd{\alpha}}} ].
  \end{gathered}
\end{equation}

Используя соотношение~\eqref{eq:k-langevin}, из соотношений~\eqref{eq:fp_coeff}
можно получить коэффициенты для уравнения Ланжевена~\eqref{eq:langevin}.

\section{Операторный подход}
\label{sec:operatorial}

Методика применения формализма вторичного квантования для неквантовых
систем (статистических, детерминированных) была рассмотрена в целом
ряде статей~\cite{Doi:1976:second_quantization,
  Doi:1976:stochastic_theory, grassberger:1980:fock-space,
  peliti:1985:path_integral_approach}.
Для записи представления чисел заполнения обычно используют нотацию
Дирака.

Переход к пространству чисел заполнения не является унитарным
преобразованием. Однако мы можем построить алгоритм перехода.

Запишем основное кинетическое уравнение~\eqref{eq:mas_eq} в
представлении чисел заполнения.  Таким образом, мы будем рассматривать
систему, не зависящую от пространственных переменных. Для простоты
рассмотрим одномерный вариант.

Обозначим в~\eqref{eq:mas_eq} через $\varphi_{n}$ вероятность
обнаружить в системе $n$ частиц:
\begin{equation}
  \label{eq:phi_n}
  \varphi_{n} := p_{n} (\varphi, t).
\end{equation}

Состояния $\varphi$ образуют векторное пространство $\mathcal{H}$.

В зависимости от структуры задачи можно ввести скалярные произведения
на основе вероятности или на основе моментов~\cite{grassberger:1980:fock-space}.
Введём скалярное произведение, исключающее
($\braket{}{}_{\text{ex}}$) и включающее ($\braket{}{}_{\text{in}}$),
считая $\ket{n}$ за базисные векторы:
\begin{gather}
  \label{eq:scalar_prod_ex}
  \braket{\varphi}{\psi}_{\text{ex}} = \sum_n n! p_{n}^{*} (\varphi) p^{n}(\psi); \\
  \label{eq:scalar_prod_in}
  \braket{\varphi}{\psi}_{\text{in}} = \sum_n \frac{1}{k!} n_{k}^{*} (\varphi)
  n^{k}(\psi).
\end{gather}
Здесь через $n_k$ обозначены факториальные моменты:
\begin{equation}
  \label{eq:factorial_moment}
  n_k (\varphi) = \expval{n(n-1) \cdots (n-k+1)} = \pdv[k]{z}
  G(z,\varphi) |_{z=1},
\end{equation}
с производящей функцией:
\begin{equation}
  \label{eq:generating_function}
  G (z, \varphi) = \sum_{n} z^{n} p_{n}(\varphi).
\end{equation}

Введём операторы рождения и уничтожения:
\begin{equation}
  \label{eq:creat+annihil}
  \begin{gathered}
    \pi \ket{n} = \ket{n+1}, \\
    a \ket{n} = n \ket{n-1}
  \end{gathered}
\end{equation}
с коммутационным соотношением\footnote{Действительно, $a\pi\ket{n} -
  \pi a \ket{n} = (n+1)\ket{n} - n \ket{n} = \ket{n}$.}:
\begin{equation}
  \label{eq:commutator}
  [a,\pi] = 1.
\end{equation}

Из~\eqref{eq:commutator} видно, что при выборе скалярного произведения
в виде~\eqref{eq:scalar_prod_ex} система описывается статистикой
Бозе--Эйнштейна.

В формализме чисел заполнения основное кинетическое уравнение
переходит в уравнение Лиувилля:
\begin{equation}
  \label{eq:liuville}
  \pdv{}{t} \ket{\varphi(t)} =  L \ket{\varphi(t)}.
\end{equation}
Оператор Лиувилля $L$ удовлетворяет соотношению:
\begin{equation}
  \label{eq:liuville=0}
  \bra{0} L = 0.
\end{equation}

Опишем диаграммную технику в рамках операторного подхода. 

\begin{figure*}
  \begin{minipage}{0.45\linewidth}
    \centering
    \includegraphics[width=0.7\linewidth]{reaction}
  \caption{Прямое взаимодействие}
    \label{fig:reaction}
  \end{minipage}
  \hfill
  \begin{minipage}{0.45\linewidth}
    \centering
    \includegraphics[width=0.7\linewidth]{reaction-rev}
    \caption{Обратное взаимодействие}
    \label{fig:reaction-rev}
  \end{minipage}
\end{figure*}

Будем записывать схемы взаимодействия в виде диаграмм. Каждой схеме
взаимодействия \eqref{eq:chemkin} или \eqref{eq:chemkin2}
соответствует пара диаграмм (рис.~\ref{fig:reaction} и
\ref{fig:reaction-rev}) для прямого и обратного взаимодействия
соответственно.
Диаграмма состоит из следующих элементов.

\begin{itemize}
\item Входящие линии (на рисунке \ref{fig:reaction} обозначено
  сплошной линией). Эти линии направлены к линии взаимодействия. Линия
  помечается количеством и типом взаимодействующих сущностей. Можно
  записывать по одной сущности на линию или группировать их.
\item Исходящие линии (на рисунке \ref{fig:reaction} обозначено
  сплошной линией). Эти линии направлены от линии взаимодействия. Линия
  помечается количеством и типом взаимодействующих сущностей. Можно
  записывать по одной сущности на линию или группировать их.
\item Линия взаимодействия (на рисунке \ref{fig:reaction} обозначена
   пунктирной линией). Направление времени обозначено стрелкой. Линия
   помечается коэффициентом интенсивности взаимодействия.
\end{itemize}

Каждой линии приписывается определённый фактор (в зависимости от
выбранного подхода). Результирующее выражение получается перемножением
этих факторов.

При применении операторного подхода с помощью диаграмм взаимодействия
мы получаем оператор Лиувилля. Каждой линии присвоим соответствующий
фактор. Результирующий членом будет получен из нормального
произведения
факторов.

\begin{figure*}
  \begin{minipage}{0.45\linewidth}
    \centering
    \includegraphics[width=0.7\linewidth]{reaction-operator}
    \caption{Прямое взаимодействие}
    \label{fig:reaction-operator}
  \end{minipage}
  \hfill
  \begin{minipage}{0.45\linewidth}
    \centering
    \includegraphics[width=0.7\linewidth]{reaction-rev-operator}
    \caption{Обратное взаимодействие}
    \label{fig:reaction-rev-operator}
  \end{minipage}
\end{figure*}

Будем использовать следующие факторы для каждого типа линий
(рис.~\ref{fig:reaction-operator}).
\begin{itemize}
\item Входящая линия. Линия соответствует выводу одной сущности из
  системы. Следовательно, ей соответствует оператор уничтожения
  $a$. Очевидно, что комбинированной линии мощности $I$ соответствует
  оператор $a^{I}$.
\item Исходящая линия. Линия соответствует появлению в системе новой
  сущности. Следовательно, ей соответствует оператор рождения
  $\pi$. Очевидно, что комбинированной линии мощности $F$ соответствует
  оператор $\pi^{F}$.
\item Линия взаимодействия. Этой линии соответствует собственно
  коэффициент интенсивности взаимодействия.
\end{itemize} 

\begin{figure*}
  \begin{minipage}{0.45\linewidth}
    \centering
    \includegraphics[width=0.7\linewidth]{reaction-operator-ext}
    \caption{Прямое взаимодействие, расширенная нотация}
    \label{fig:reaction-operator-ext}
  \end{minipage}
  \hfill
  \begin{minipage}{0.45\linewidth}
    \centering
    \includegraphics[width=0.7\linewidth]{reaction-rev-operator-ext}
    \caption{Обратное взаимодействие, расширенная нотация}
    \label{fig:reaction-rev-operator-ext}
  \end{minipage}
\end{figure*}

То есть, для диаграммы~\ref{fig:reaction-operator} мы получим фактор
$\tensor*[^{+}]{k}{} \pi^{F} a^{I}$. Однако при этом нарушается
соотношение~\eqref{eq:liuville=0}. Для исправления этого положения мы
должны вычесть количество сущностей, вступивших во взаимодействие,
помноженное на интенсивность взаимодействия. Тогда получим следующий
член оператора Лиувилля:
\begin{equation}
  \label{eq:diagram-factor}
  \tensor*[^{+}]{k}{} \pi^{F} a^{I}  - \tensor*[^{+}]{k}{} 
  \pi^{I} a^{I} =
  \tensor*[^{+}]{k}{} 
  \qty(\pi^{F}  - \pi^{I}) a^{I}.
\end{equation}

Для обратных взаимодействий (рис.~\ref{fig:reaction-rev-operator})
используются эти же правила.

Для учёта дополнительного фактора~\eqref{eq:diagram-factor} будем
использовать расширенные диаграммы
(см. рис.~\ref{fig:reaction-operator-ext} и
\ref{fig:reaction-rev-operator-ext}). Здесь из нормального
произведения числителей вычитается нормальное произведение знаменателей.

Таким образом, схеме~\eqref{eq:chemkin} соответствует оператор Лиувилля:
\begin{equation}
  \label{eq:liuville:chemkin}
  L = \sum_{\crd{\alpha},\crd{i}} \biggl[
  \tensor*[^{+}]{k}{_{\crd{\alpha}}}
  \qty(
  (\pi_{\crd{i}})^{F^{\crd{i}\crd{\alpha}}} -
  (\pi_{\crd{i}})^{I^{\crd{i}\crd{\alpha}}} 
  ) (a_{\crd{i}})^{I^{\crd{i}\crd{\alpha}}}
  + 
  \tensor*[^{-}]{k}{_{\crd{\alpha}}}
  \qty(
  (\pi_{\crd{i}})^{I^{\crd{i}\crd{\alpha}}} -
  (\pi_{\crd{i}})^{F^{\crd{i}\crd{\alpha}}} 
  ) (a_{\crd{i}})^{F^{\crd{i}\crd{\alpha}}}
  \biggr].
\end{equation}

\section{Модель Ферхюльста}
\label{sec:model}

В качестве демонстрации метода рассмотрим модель
Ферхюльста~\cite{verhulst:1838, Feller:1939:acta_biotheoretica,
  feller:1949:theory_stochastic_processes}, описывающую ограниченный
рост. Привлекательность этой модели в том, что она одномерна и
  нелинейна. Изначально эта модель описывается следующим
дифференциальным уравнением:
\begin{equation}
    \dv{\varphi}{t}= \lambda \varphi - \beta \varphi - \gamma \varphi^{2},
\end{equation}
здесь $\lambda$~--- коэффициент интенсивности размножения, $\beta$~---
коэффициент интенсивности вымирания, $\gamma$~--- коэффициент
интенсивности уменьшения популяции (обычно рассматривается
соперничество особей). Здесь мы оставляем те же обозначения,
  что и в исходной модели~\cite{verhulst:1838}.

Построим стохастический вариант данной модели. Запишем схемы
взаимодействия:
\begin{equation}
  \label{eq:verhulst:chemkin}
  \begin{gathered}
    \varphi \overset{\lambda}{\underset{\gamma}{\rightleftharpoons}} 2\varphi ,\\
    \varphi \xrightarrow{\beta} 0.
  \end{gathered}
\end{equation}

Схемам взаимодействия~\eqref{eq:verhulst:chemkin} соответствуют
диаграммы~\ref{fig:verhulst-1}, \ref{fig:verhulst-1-rev} и \ref{fig:verhulst-2}.

Первое соотношение означает, что индивидуум, который съедает единицу
пищи, немедленно репродуцируется, в обратную сторону~--
соперничество между индивидуумами. Второе~---  смерть индивидуума.

\begin{figure*}
  \begin{minipage}{0.3\linewidth}
    \centering
    \includegraphics[width=\linewidth]{verhulst-1}
  \caption{Первое прямое взаимодействие}
\label{fig:verhulst-1}
  \end{minipage}
  \hfill
  \begin{minipage}{0.3\linewidth}
    \centering
    \includegraphics[width=\linewidth]{verhulst-1-rev}
    \caption{Первое обратное взаимодействие}
    \label{fig:verhulst-1-rev}
  \end{minipage}
  \hfill
  \begin{minipage}{0.3\linewidth}
    \centering
    \includegraphics[width=0.8\linewidth]{verhulst-2}
    \caption{Второе взаимодействие}
    \label{fig:verhulst-2}
  \end{minipage}
\end{figure*}

\subsection{Комбинаторный подход}

Определим интенсивности переходов для модели Ферхюльста:
\begin{equation}
  \begin{gathered}
    \tensor*[^{+}]{s}{_{1}} (\varphi) = \lambda \varphi, \\
    \tensor*[^{-}]{s}{_{1}} (\varphi) = \gamma \varphi (\varphi - 1), \\
    \tensor*[^{+}]{s}{_{2}} (\varphi) = \beta \varphi.
  \end{gathered}
  \quad
  \begin{gathered}
    \tensor*[^{+}]{s}{_{1}} (\varphi -1) = \lambda (\varphi -1), \\
    \tensor*[^{-}]{s}{_{1}} (\varphi -1) = \gamma (\varphi -1) (\varphi - 2), \\
    \tensor*[^{+}]{s}{_{2}} (\varphi -1) = \beta (\varphi -1).
  \end{gathered}
  \quad
  \begin{gathered}
    \tensor*[^{+}]{s}{_{1}} (\varphi +1) = \lambda (\varphi +1), \\
    \tensor*[^{-}]{s}{_{1}} (\varphi +1) = \gamma (\varphi +1) \varphi, \\
    \tensor*[^{+}]{s}{_{2}} (\varphi +1) = \beta (\varphi +1).
  \end{gathered}
\end{equation}

\begin{equation}
  r^1 = 1, \qquad
  r^2 = -1.
\end{equation}

Тогда (на основании~\eqref{eq:one-step:master:full}) основное кинетическое
уравнение примет следующий вид:
\begin{equation}
  \label{eq:master}
  \pdv{p (\varphi,t)}{t} =  
  - \qty[\lambda \varphi + \beta \varphi +
  \gamma \varphi (\varphi -1)] 
  p(\varphi,t) 
  +
  \qty[\beta (\varphi + 1)
  +  \gamma (\varphi + 1) \varphi] 
  p(\varphi + 1,t) + \lambda (\varphi - 1) p (\varphi - 1,t).
\end{equation}

Или же, записывая для $\varphi=n$ (см.~\eqref{eq:mas_eq}):
\begin{multline}
  \label{eq:verhulst:master:n}
  \pdv{p_{n}(t)}{t} := \pdv{p (\varphi,t)}{t}\eval_{\varphi=n} 
  = {} \\ {} =
  - \qty[\lambda n + \beta n +
  \gamma n(n-1)] 
  p_{n}(t) 
  +
  \qty[\beta (n + 1) + \gamma (n + 1)n] 
  p_{n+1}(t) + 
  \lambda (n - 1) p_{n-1} (t).
\end{multline}

\subsection{Операторный подход}

Схемам взаимодействия~\eqref{eq:verhulst:chemkin} в операторном подходе соответствуют
диаграммы~\ref{fig:verhulst-1-operator}, \ref{fig:verhulst-1-rev-operator} и \ref{fig:verhulst-2-operator}.

На основании~\eqref{eq:verhulst:chemkin} и~\eqref{eq:liuville:chemkin}
получаем оператор Лиувилля:
\begin{multline}
  \label{eq:verhulst:liuville}
  L = \lambda (\pi^2 -\pi) a + \gamma (\pi - \pi^2) a^2 + 
  \beta (1 - \pi) a 
  = {} \\ {} =
  \lambda \qty((a^{\dagger})^2 - a^{\dagger}) a + 
  \gamma \qty(a^{\dagger} - (a^{\dagger})^2) a^2 + 
  \beta \qty(1 - a^{\dagger}) a 
  = {} \\ {} =
  \lambda \qty(a^{\dagger} - 1)a^{\dagger} a +
  \beta \qty(1 - a^{\dagger}) a + 
  \gamma \qty(1 - a^{\dagger})a^{\dagger} a^2.
\end{multline}

\begin{figure*}
  \begin{minipage}{0.3\linewidth}
    \centering
    \includegraphics[width=\linewidth]{verhulst-1-operator}
  \caption{Первое прямое взаимодействие}
\label{fig:verhulst-1-operator}
  \end{minipage}
  \hfill
  \begin{minipage}{0.3\linewidth}
    \centering
    \includegraphics[width=\linewidth]{verhulst-1-rev-operator}
    \caption{Первое обратное взаимодействие}
    \label{fig:verhulst-1-rev-operator}
  \end{minipage}
  \hfill
  \begin{minipage}{0.3\linewidth}
    \centering
    \includegraphics[width=0.8\linewidth]{verhulst-2-operator}
    \caption{Второе взаимодействие}
    \label{fig:verhulst-2-operator}
  \end{minipage}
\end{figure*}

Запишем основное кинетическое уравнение через оператор Лиувилля:
\begin{multline}
  \label{eq:verhulst:master2liuville}
  \pdv{p_{n}(t)}{t} = \frac{1}{n!} \bra{n}L\ket{\varphi} 
  =  {} \\ {} =
  \frac{1}{n!} \bra{n}
  - \qty[ 
  \lambda a^{\dagger} a +
  \beta a^{\dagger} a +
  \gamma a^{\dagger}a^{\dagger} a a
  ]
  +
  \qty[
  \beta a + 
  \gamma a^{\dagger} a a
  ]
  +
  \lambda a^{\dagger}a^{\dagger} a 
  \ket{\varphi}
  = {} \\ {} =
  - \qty[\lambda n + \beta n +
  \gamma n(n-1)] 
  \bra{n}\ket{\varphi}
  + {} \\ {} +
  \qty[\beta (n + 1) + \gamma (n + 1)n] 
  \bra{n+1}\ket{\varphi} + 
  \lambda (n - 1) \bra{n-1}\ket{\varphi}
  = {} \\ {} =
  - \qty[\lambda n + \beta n +
  \gamma n(n-1)] 
  p_{n}(t) 
  + {} \\ {} +
  \qty[\beta (n + 1) + \gamma (n + 1)n] 
  p_{n+1}(t) + 
  \lambda (n - 1) p_{n-1} (t).
\end{multline}

Результат~\eqref{eq:verhulst:master2liuville} полностью совпадает с
формулой~\eqref{eq:verhulst:master:n}, полученной комбинаторным методом.

\section{Заключение}
\label{sec:conclusion}

Описанные методы стохастизации позволяют получать конкретный вид как
оператора Лиувилля, так и самосогласованных стохастических
дифференциальных уравнений в форме Ланжевена и Фоккера--Планка. Авторы
надеются, что предложенная диаграммная техника упростит получение
оператора Лиувилля в представлении чисел заполнения.

\def\acknowledgmentsname{Благодарности}

\begin{acknowledgments}

Публикация подготовлена при поддержке Программы РУДН «5-100»
и при финансовой поддержке РФФИ в рамках научных проектов
\textnumero~16-07-00556, 18-07-00567.

\end{acknowledgments}

  \ifthenelse{\boolean{xetex}}
  {\bibliographystyle{ugost2008l}}
  {\bibliographystyle{gost2008l}}

\bibliography{bib/gen-sdu/cite}